 \documentclass[12pt,prb,aps,preprint]{revtex4}
\begin{document}
\def\ie{{\it i.\ e.\ }}
\def\ni {\noindent}
\title{On the critical dipole moment in one-dimension}
\author{Benjam\'{\i}n Jaramillo}\email{benraz@licifug.ugto.mx}
\affiliation{ Instituto de F\'{\i}sica, Universidad de Guanajuato, Loma del Bosque 103, Fracc.\ Lomas del Campestre, C P 37150 Le\'on, Guanajuato, M\'exico}
\author{  H. N. N\'u\~nez-Y\'epez}
\email{nyhn@xanum.uam.mx}
\affiliation{ Departamento de F\'{\i}sica, Universidad Aut\'onoma Metropolitana, Unidad Iztapalapa, Apartado Postal 55-534, Iztapalapa CP 09340 D. F. M\'exico}
\author{ A. L. Salas-Brito\footnote{Author to whom all correspondence should be addressed.}}\email{asb@correo.azc.uam.mx}
\affiliation{ Laboratorio de Sistemas Din\'amicos, Universidad Aut\'onoma Metropolitana, Unidad Azcapotzalco,  Aparta\-do Pos\-tal 21-267, C P 04000, Coyoac\'an D. F. M\'e\-xico}
 \maketitle
In a recent paper Connolly and Griffiths (CG) calculated the minimum magnitude, $p_{{}_{\hbox{crit}}}$, an electric dipole must have for the dipolar system to support bound states. The difficulties  one may encounter in calculating the critical dipole moment  in one, two, and three dimensions  are also explained. In  the one-dimensional case  there may not be a critical dipole moment since ---they argued--- in one dimension the Coulomb  ground state has infinite binding energy.\cite{kcjg07}

Contrary to what CG  hint, in this work we are able to estimate $p_{\hbox{crit}}$ using the spectrum of the one dimensional hydrogen atom.\cite{ejp87,fisher95,tsutsui03} Then, by using  the point dipole potential  in one dimension,\cite{kcjg07}

\begin{equation}\label{0}
V_{\hbox{\tiny pd}}(x)=p\frac{\kappa}{x|x|}
\end{equation}

\noindent where $\kappa\equiv q/(4\pi\epsilon_0)$, we compute the exact value of the critical dipole moment. Therefore, contrary to the claims about its possible non existence, we  exhibit that $p_{\hbox{crit}}$ do indeed exists. Notice that the $p_{\hbox{crit}}$ calculated for the point dipole model is also the critical dipole moment for a physical dipole with charges separated a certain distance  since its value is independent of the separation.\cite{kcjg07,levy67}

For the calculation, CG considered at first the 1D Coulomb  potential, $V_0(x)=-{\lambda}/|x|$, where $\lambda=Q\kappa$, stating, following Loudon,\cite{loud59} that its  ground state  has an infinite binding energy. Consequently, they concluded that such feature may prevent the existence of a critical dipole moment in one dimension. Thus, if the ground state for the 1D hydrogen atom were finite, a value for the critical dipole moment could be estimated as we do below. 

CG made plausible the existence of the infinite binding energy ground state by exhibiting the results of a numerical computation with  the  regularized potential 
\begin{equation}
V_{\varepsilon}(x)=\left\{\begin{array}{ll}-\lambda/\varepsilon &\textrm{if $|x|\le\varepsilon$}\\
                                               -\lambda/|x| &\textrm{if $|x|\ge\varepsilon$},
                              \end{array}\right.  
\end{equation}

\noindent where $\varepsilon$ is a regularizing cut-off, showing that the absolute value of the ground state energy increases when $\varepsilon$ becomes smaller. We question such argument because, in spite of the numerics, no  state of infinite binding energy exist for the Hamiltonian \cite{ejp87,fisher95,tsutsui03}
\begin{equation}\label{h0}
\widehat{H}_0=-\frac{\hbar^2}{2m} \frac{d ^2}{d x^2}+ V_0(x) 
\end{equation}
\noindent of the one-dimensional hydrogen atom. 

 To obtain a correct solution of this problem, one would need first to determine the  self-adjoint extensions \cite{adjoint} of  (\ref{h0}). The  energy spectrum is shown then to contain no state of infinite energy. \cite{fisher95,tsutsui03}
But, although not using explicitly self-adjoint extensions, both Andrews\cite{acan66} and N\'u\~nez-Y\'epez and  Salas-Brito\cite{nyetal87} had previously shown that the  infinite binding energy ground state, predicted by Loudon, would vanish in the limit $\varepsilon \to 0$. Boya {\it et al.\ } \cite{boya89} have also concluded that such state needs to be discarded.    Thus, analysing  what happens to the ground state energy  of the  cut-off potential as $\varepsilon$ gets smaller and smaller,  gives no information about  the ground state energy of the \begin{math} -1/ \vert x \vert\end{math} potential. 
In fact, the energy spectrum of the one-dimensional hydrogen atom, requiring  the wave functions to vanish at the origin,\cite{dirichlet} is given by the Balmer formula, $E_{n}=-{q\;Q}/(8 \pi \epsilon_{0} a_B n^2)$, $n=1,2,3,\dots$ where $Q$ is the charge of the nucleus, $a_B={4 \pi \epsilon_{0} \hbar^2}/(q\,Q m)$ is the Bohr radius, $q$ is the electronic charge,  and $m$ is the mass of the electron, exactly as in the three dimensional case.\cite{ejp87,boya89,pogosyan89,fisher95,tsutsui03} 
The energy of the ground state is thus  $-{q\,Q}/(8 \pi \epsilon_{0} a_B)$.

Moreover, once the infinite energy state is discarded, we may estimate the distance, $d$, at which the ionization of a 
one-dimensional atom occurs due to the presence of another charge,\cite{kcjg07} $Q$, and therefore  we may get a rough estimate  of the critical  dipole moment equating the Coulomb repulsion energy $E_{re}=q\,Q/(4\pi\epsilon_0 d)$ with the energy of the ground state, $E_1$, we  obtain 
\begin{equation}
p_{\hbox{crit}}^{\hbox{(est)}}=Q\,d=8\pi\epsilon_0\frac{\hbar^2}{qm}.
\end{equation}
\ni As can be seen on comparing with (\ref{pcrit}), this rough estimate is  sixteen times larger than the exact value. But, its real importance is that it suggests that  a critical value for the dipole moment in one-dimension may exist. Our next task is to compute $p_{\hbox{crit}}$ exactly and, by doing so, to establish its existence  beyond any doubt. 

The Schr\"odinger equation describing an electron interacting with a point dipole is\cite{kcjg07}  
 \begin{equation}\label{5} 
-\frac{\hbar^{2}}{2m} \frac{d^{2} \Psi}{dx^{2}} + p\frac{\kappa}{x|x|} \Psi(x)=E \Psi(x).  
\end{equation}

To solve (\ref{5}) notice that for $x>0$ the potential is repulsive and no bound states exist, hence $\Psi(x>0)=0$.\cite{aejg06} There is as an impenetrable barrier at $x=0$  so any  bound particle has zero probability of crossing to the right side.\cite{garba04}
The solutions must then approach zero as $x \to 0$ from the left. Introducing $y=-x$ in Eq.\ (\ref{5}), we get 
\begin{equation}\label{6} 
\label{schro} -\frac{d^{2} \Psi}{dy^{2}} - \frac{\alpha}{y^{2}} \Psi(y)= -\xi \Psi(y)  
\end{equation}
\noindent where $ \alpha \equiv {2 m p q}/{(4 \pi \epsilon_{0} \hbar^{2})} > 0$ and $ \xi \equiv -{2mE}/{\hbar^{2}}$.  We write $\Psi(y)$ as the power series 
\begin{equation}\label{suma}
\Psi(y)= \sum_{j=0}^{\infty} a_{j} y^{j + \nu},\quad (\hbox{we assume } a_0\neq 0)
\end{equation}
\noindent then substituting (\ref{suma}) into (\ref{6}), it becomes
\begin{equation} \label{7}
\sum_{j=0}^{\infty}a_{j}y^{j+\nu-2}[(j+\nu)(j+\nu-1)+\alpha]=\xi\sum_{j=0}^{\infty}a_{j}y^{j+\nu},
\end{equation}
\noindent leading to the  relationships
\begin{eqnarray} \label{a}
[\nu(\nu-1)+\alpha]a_{0}&=&0,\cr
[\nu(\nu+1)+\alpha]a_{1}&=&0,\cr
[(\nu+j+2)(\nu+j+1)+\alpha]a_{j+2}&=&\xi a_{j}.
\end{eqnarray}

\noindent From (\ref{a}) we conclude that all the odd-term coefficients vanish and we find two possible values for the leading exponent,   $\nu_{\pm}=(1\pm\sqrt{1-4\alpha})/2$.  Therefore, there are two independent solutions, $\Psi_{\pm}$, behaving near $y=0$  as\cite{kcjg07}
\begin{equation}
\Psi_{\pm}(y)\sim a_0 y^{1/2} e^{\pm \sqrt{1/4-\alpha} \ln y}.
\end{equation}
\noindent These solutions are finite at the origin and square integrable only if  $ \sqrt{1/4-\alpha}$ is imaginary, thus  $\alpha\geq 1/4$ for bound states to exist.\cite{calculo,kggr93,aejg06} Using this result we obtain the critical value  of the point dipole in one dimension
\begin{equation}\label{pcrit}
p_{\hbox{crit}} = \frac{\pi\epsilon_{0}}{2}\frac{\hbar^2}{ q m}.
\end{equation}
\noindent  We emphasize that a physical dipole has exactly the same critical value as the point dipole. Therefore, a one-dimensional dipole does not always support bound states; for that to happen the dipole moment must be larger or at least equal than $p_{\hbox{crit}}$. We must point out that, even  in such a case, there can  only be one bound state.\cite{kcjg07, levy67, kggr93}

{\bf Conclusion:} In one-dimension there is a critical dipole moment $p_{\hbox{crit}}= \pi\epsilon_{0}\hbar^2/2q m= 1.052\times 10^{-30} \hbox{ C}\cdot\hbox{m}$. Any one-dimensional system with an electric dipole moment smaller than $p_{\hbox{crit}}$ cannot support bound states.

 {\bf Acknowledgements.}\par BJ wants  to acknowledge the support of the Academia Mexicana de Ciencias through the Verano de la Ciencia program. This work was partially supported by a PAPIIT-UNAM research grant (IN115406-3). We also acknowledge with thanks the comments of Roberto Sussman.


\begin{thebibliography}{000}
\bibitem{calculo} This calculation is done for the radial part of the three dimensional point dipole in Ref.\ [\cite{kcjg07}]. For the the one-dimensional $-a/x^{2}$ potential see Ref.\ [\cite{aejg06}].
\end{thebibliography}
\end{document}